\documentclass[12pt]{article}

\usepackage{latexsym}

\begin{document}

\title{Plane-symmetric inhomogeneous Brans-Dicke cosmology
with an equation of state $p=\gamma \rho $}

\author{
     Stoytcho S. Yazadjiev \thanks{E-mail: yazad@phys.uni-sofia.bg}\\
{\footnotesize  Department of Theoretical Physics,
                Faculty of Physics, Sofia University,}\\
{\footnotesize  5 James Bourchier Boulevard, Sofia~1164, Bulgaria }\\
}

\date{}

\maketitle

\begin{abstract}
We present a new exact solution in Brans-Dicke theory. The
solution describes inhomogeneous plane-symmetric perfect fluid
cosmological model with an equation of state $p=\gamma \rho$.
Some main properties of the solution are discussed.
\end{abstract}


\sloppy
\renewcommand{\baselinestretch}{1.3} %
\newcommand{\sla}[1]{{\hspace{1pt}/\!\!\!\hspace{-.5pt}#1\,\,\,}\!\!}
\newcommand{\db}{\,\,{\bar {}\!\!d}\!\,\hspace{0.5pt}}
\newcommand{\partb}{\,\,{\bar {}\!\!\!\partial}\!\,\hspace{0.5pt}}
\newcommand{\dsla}{\partb}
\newcommand{\eql}{e _{q \leftarrow x}}
\newcommand{\eqr}{e _{q \rightarrow x}}
\newcommand{\ite}{\int^{t}_{t_1}}
\newcommand{\itz}{\int^{t_2}_{t_1}}
\newcommand{\itd}{\int^{t_2}_{t}}
\newcommand{\lfrac}[2]{{#1}/{#2}}
\newcommand{\dV}{d^4V\!\!ol}
\newcommand{\ben}{\begin{eqnarray}}
\newcommand{\een}{\end{eqnarray}}
\newcommand{\la}{\label}


\section{Introduction}

String theory and higher dimensional unifying theories, in their
low energy limit, predict the existence of a scalar partner of
the tensor graviton - the dilaton. So obtained  scalar-tensor
theories are considered as the most natural generalization of
general relativity \cite{DP}-\cite{GAS}. A lot of effort has been
devoted to the finding and analysis of exact scalar-tensor
solutions in order to understand more deeply the physics behind
these theories, in particular their relevance to cosmology and
astrophysics \cite{GV}-\cite{Y1}.

The scalar-tensor gravity equations are much more complicated than
the Einstein equations and their solving in the presence of a
source is a very difficult task. That is why one should assume
some simplifications  in order to solve the scalar tensor
equations. In this way many homogeneous cosmological solutions
with a perfect fluid have been obtained. Some inhomogeneous
scalar-tensor cosmologies have also been found and a method  for
generating general classes of exact scalar-tensor solutions with
a stiff perfect fluid has been given \cite{Y2}, \cite{Y3}.

However, the known exact scalar-tensor solutions cover only a
small part of the  physical content of the scalar-tensor
equations. The search of new exact solutions is therefore
necessary if further progress is to be made in understanding the
scalar-tensor theories. The purpose of this paper is to present a
new exact plane-symmetric solution in Brans-Dicke theory with a
prefect fluid satisfying the equation of state $p=\gamma \rho$.
The found solution can be interpreted as an inhomogeneous
cosmology. The study of inhomogeneous cosmologies is necessary
because, as well known, the present universe is not exactly
spacialy homogeneous. Although the homogeneous  models are good
approximations of the present universe there is no reasons  to
assume that such a regular expansion is also suitable for the
description of the early universe. Moreover, it has been shown
that the existence of large inhomogeneities in the universe does
not necessarily lead to an observable trace left over the
spectrum of CMB \cite{RT}-\cite{AFS}. It has also been
demonstrated the existence of homogeneous but highly  anisotropic
cosmological models whose CMB  is exactly isotropic
\cite{LNW},\cite{CMM}.

In the light of these results the study of inhomogeneous and
anisotropic cosmological models is even imperative.

\section{Exact solution with a plane symmetry}

The general form of the extended gravitational action in
scalar-tensor theories is

\begin{eqnarray} \label{JFA}
S = {1\over 16\pi G_{*}} \int d^4x \sqrt{-{\tilde
g}}\left({F(\Phi)\tilde R} - Z(\Phi){\tilde
g}^{\mu\nu}\partial_{\mu}\Phi
\partial_{\nu}\Phi  \right. \nonumber  \\ \left. -2 U(\Phi) \right) +
S_{m}\left[\Psi_{m};{\tilde g}_{\mu\nu}\right] .
\end{eqnarray}

Here, $G_{*}$ is the bare gravitational constant, ${\tilde R}$ is
the Ricci scalar curvature with respect to the space-time metric
${\tilde g}_{\mu\nu}$. The dynamics of the scalar field $\Phi$
depends on the functions $F(\Phi)$, $Z(\Phi)$ and $U(\Phi)$. In
order for the gravitons  to carry positive energy the function
$F(\Phi)$ must be positive. The nonnegativity of the energy of
the dilaton field requires that $2F(\Phi)Z(\Phi) +
3[dF(\Phi)/d\Phi]^2 \ge 0$. The action of matter depends on the
material fields $\Psi_{m}$ and the space-time metric ${\tilde
g}_{\mu\nu}$. It should be noted that the stringy generated
scalar-tensor theories, in general, admit the direct interaction
between the matter fields and the dilaton in the Jordan (string)
frame \cite{DP}, \cite{GV}. Here we consider the phenomenological
case when the matter action does not involve the dilaton field in
order for the weak equivalence principle to be satisfied.

However, it is much more convenient from a mathematical point of
view to analyze the scalar-tensor theories with respect to the
conformally  related Einstein frame  given by the metric:

\begin{equation}\label {CONF1}
g_{\mu\nu} = F(\Phi){\tilde g}_{\mu\nu} .
\end{equation}

Further, let us introduce the scalar field $\varphi$ (the so
called dilaton) via the equation

\begin{equation}\label {CONF2}
\left(d\varphi \over d\Phi \right)^2 = {3\over
4}\left({d\ln(F(\Phi))\over d\Phi } \right)^2 + {Z(\Phi)\over 2
F(\Phi)}
\end{equation}

 and define

\begin{equation}\label {CONF3}
{\cal A}(\varphi) = F^{-1/2}(\Phi) \,\,\, ,\nonumber \\
2V(\varphi) = U(\Phi)F^{-2}(\Phi) .
\end{equation}

In the Einstein frame action (\ref{JFA}) takes the form

\begin{eqnarray}
S= {1\over 16\pi G_{*}}\int d^4x \sqrt{-g} \left(R -
2g^{\mu\nu}\partial_{\mu}\varphi \partial_{\nu}\varphi -
4V(\varphi)\right) \nonumber \\ + S_{m}[\Psi_{m}; {\cal
A}^{2}(\varphi)g_{\mu\nu}]
\end{eqnarray}

where $R$ is the Ricci scalar curvature with respect to the
Einstein metric $g_{\mu\nu}$.

The Einstein frame field equations then are

\begin{eqnarray} \label{EFFE}
R_{\mu\nu} - {1\over 2}g_{\mu\nu}R = 8\pi G_{*} T_{\mu\nu}
 + 2\partial_{\mu}\varphi \partial_{\nu}\varphi \nonumber \\  -
g_{\mu\nu}g^{\alpha\beta}\partial_{\alpha}\varphi
\partial_{\beta}\varphi -2V(\varphi)g_{\mu\nu}  \,\,\, ,\nonumber
\end{eqnarray}

\begin{eqnarray}
 \nabla^{\mu}\nabla_{\mu}\varphi = - 4\pi G_{*} \alpha (\varphi)T
+ {dV(\varphi)\over d\varphi} \,\,\, ,
\end{eqnarray}

\begin{eqnarray}
\nabla_{\mu}T^{\mu}_{\nu} = \alpha
(\varphi)T\partial_{\nu}\varphi \,\,\, . \nonumber
 \end{eqnarray}

Here $\alpha(\varphi)= {d\ln({\cal  A}(\varphi))/ d\varphi}$ and
the Einstein frame energy-momentum tensor $T_{\mu\nu}$  is
related to the Jordan frame one ${\tilde T}_{\mu\nu}$ via
$T_{\mu\nu}= {\cal A}^2(\varphi){\tilde T}_{\mu\nu}$. In the case
of a perfect fluid one has

\begin{eqnarray}\label{DPTEJF}
\rho &=&{\cal A}^4(\varphi){\tilde \rho}, \nonumber \\
p&=&{\cal A}^4(\varphi){\tilde p},  \\
u_{\mu}&=& {\cal A}^{-1}(\varphi){\tilde u}_{\mu} . \nonumber
\end{eqnarray}

The mathematical complexity  makes it very difficult to find
inhomogeneous solutions  of the system  (\ref{EFFE}) even for
simplified models. In fact, when
$g^{\mu\nu}\partial_{\mu}\varphi\partial_{\mu}\varphi < 0$ the
system
 (\ref{EFFE}) describes two interacting fluids in general relativity. As 
far as we are aware, there is no scalar-tensor inhomogeneous cosmological 
solution with an equation
of state $p=\gamma \rho$. In the particular case $\gamma=1/3$
($T=0$), we have noninteracting  fluids and some inhomogeneous
solutions can be found \cite{Y4}.

In this paper we consider Brans-Dicke theory with $V(\varphi)=0$.
This theory is described by the functions $F(\Phi)=\Phi$ and
$Z(\Phi)=\omega/\Phi$ corresponding to ${\cal A}(\varphi)=
exp(\alpha \varphi)$ where $\alpha = 1/\sqrt{3 + 2\omega}$.

We have succeeded in finding the following solution of the field
equations (\ref{EFFE}) for a perfect fluid with an equation of
state $p=\gamma \rho$ (${\tilde p}=\gamma {\tilde \rho}$ ) and
$0<\gamma <1$:

\begin{eqnarray}
ds^2 = \cosh^{{1 + 3\gamma}\over \lambda (1-\gamma)}\left(\lambda
a x\right)[-d\tau^2 + \mu^2 a^2\tau^2 dx^2 ] \nonumber \\ +
\left(\mu a\tau\right)^{1\over \mu} \cosh^{-{1\over
\lambda}}\left(\lambda a x\right)[dy^2 + dz^2],
\end{eqnarray}

\begin{eqnarray}
8\pi G_{*} \rho = {1+ \lambda  \over \mu^2 (1-\gamma)} {1\over
\tau^2  \cosh^{2 + {{1 + 3\gamma}\over \lambda
(1-\gamma)}}\left(\lambda a x\right)},
\end{eqnarray}

\begin{eqnarray}
u_{\mu} = - \cosh^{{1 + 3\gamma}\over 2 \lambda
(1-\gamma)}\left(\lambda a x\right) \delta^{0}_{\mu},
\end{eqnarray}

\begin{eqnarray}
\varphi = {1\over 2}\alpha {3\gamma -1\over 1-\gamma }
\left({1\over \mu} \ln\left(\mu a\tau \right)  - {1\over
\lambda}\ln\left(\cosh(\lambda a x) \right) \right),
\end{eqnarray}

where $\mu=\mu(\gamma)$  and $\lambda=\lambda(\gamma)$ are given
by

\begin{eqnarray}
2\mu(\gamma) &=&  {2(1-\gamma^2 ) + \alpha^2 (3\gamma -1)^2 \over 
(1-\gamma)^2}, \\
\lambda(\gamma) &=& { (1 + 3\gamma)(1-\gamma)  + \alpha^2
(3\gamma -1) ^2 \over 4\gamma (1-\gamma) } .
\end{eqnarray}

The solution depends on one parameter $a$ which satisfies $a>0$.
The range of the coordinates is

\begin{equation}
0< \tau < \infty, \,\,\, -\infty < x,y,z < \infty .
\end{equation}

The spacetime described by the solution is plane-symmetric
\cite{KSHMC}. It has a three dimentional group of local
isometries\footnote{Obviously, there is also a discrete isometry
given by $x\rightarrow - x$.} acting on two-dimensional orbits and
generated by the Killing vectors:

\begin{equation}
{\cal K}_{1} = {\partial\over \partial y}, \,\,\, {\cal K}_{2} =
{\partial\over \partial z}, \,\,\, {\cal K}_{3} = y{\partial\over
\partial z} - z{\partial\over \partial y}.
\end{equation}

Therefore, the found solution can be interpreted as an
inhomogeneous cosmology.

Let us note that in the particular case $\gamma=1/3$ we obtain
pure general relativistic solution with a trivial dilaton field
$\varphi=0$.

As $\tau\to 0$ the curvature invariants, energy density, pressure
and the dilaton field diverge which corresponds to a big-bang
singularity. The Einstein frame  perfect fluid expansion,
acceleration and shear calculated in the natural orthonormal
tetrad

\begin{eqnarray}
e^{0} = \mid g_{00}\mid^{1/2} dt ,\,\,\,\, e^{1} = g_{11}^{1/2}
dx, \,\,\,\, e^{2} = g_{22}^{1/2} dy, \,\,\,\, e^{3} =
g_{33}^{1/2} dz,
\end{eqnarray}

 are the following:

\begin{eqnarray}
\theta &=& (\mu +1) {a\over g^{1/2}_{11}}, \\
a_{1} &=& {1+ 3\gamma \over 2(1-\gamma)} a\, {\tanh(\lambda a x) \over 
g^{1/2}_{11}}, \\
\sigma_{11} &=& {2\mu - 1\over 3} {a\over g^{1/2}_{11}},  \\
\sigma_{22} &=& \sigma_{33} = - {1\over 2} \sigma_{11}.
\end{eqnarray}

It is seen that the fluid is everywhere expanding in the Einstein
frame. It is also interesting to note that, near the singularity,
the scalar fluid is not dominant ($\rho_{\varphi}\sim 1/\tau^2 $)
due to the interaction between the fluids.

The Jordan (string) frame solution can be obtained by using the
inverse of transformations (\ref{CONF1}),
(\ref{CONF2}),(\ref{CONF3}) and (\ref{DPTEJF}).

The properties of the Jordan frame solution depend on the value
of the parameter $\alpha$. In the physically interesting case
when $\alpha \le 1$ (i.e. when $\omega$ is equal to or greater
than the stringy value $\omega=-1$ ) the Jordan frame solution
has  qualitatively the same properties as the Einstein frame
solution.

With regard to the kinematical quantities in the Jordan frame,
the only non-vanishing components of expansion, acceleration and
shear of the fluid are given by:

\begin{eqnarray}
{\tilde \theta} &=& {2(1-\gamma) + \alpha^2 (3\gamma-1) \over
(1-\gamma)^2 } \,
{a\over {\tilde g}^{1/2}_{11}   } ,\\
{\tilde a}_{1} &=& {(1 + 3\gamma) + \alpha^2 (1-3\gamma) \over
2(1-\gamma)}a\,
{\tanh(\lambda a x)\over {\tilde g}^{1/2}_{11}   } , \\
{\tilde \sigma}_{11} &=& {2\mu -1\over 3}\, {a\over {\tilde g}^{1/2}_{11}   
} , \\
{\tilde \sigma}_{22} &=& {\tilde \sigma}_{33} =-{1\over 2}
{\tilde \sigma}_{11}.
\end{eqnarray}

All quantities are calculated in the natural orthonormal tetrad:

\begin{eqnarray}
{\tilde e}^{0} = \mid{\tilde g}_{00}\mid^{1/2} dt ,\,\,\,\,
{\tilde e}^{1} = {\tilde g}_{11}^{1/2} dx, \,\,\,\, {\tilde e}^{2}
= {\tilde g}_{22}^{1/2} dy, \,\,\,\, {\tilde e}^{3} = {\tilde
g}_{33}^{1/2} dz.
\end{eqnarray}

To the best of our knowledge the presented exact solution is the
first example of an inhomogeneous perfect fluid  scalar-tensor
cosmology with  an equation of state different form that of the
stiff matter.

\bigskip

\bigskip

 {\em \bf Acknowledgments}

\bigskip
This work was partially supported by Sofia University Grant
No.3429/2002-03.

\end{document}